\documentclass{aa}  
\usepackage{graphicx,amssymb,amsmath}
\usepackage{natbib}
\usepackage{verbatim} 
\usepackage{txfonts}
\def\spitzer{{\it Spitzer~\/}}

\def\lsun{\hbox{$\rm ~L_{\odot}$}}
\def\msun{\hbox{$\rm ~M_{\odot}$}}

\def\H0{{\rm ~km~s$^{-1}$~Mpc$^{-1}$}}

\def\ergsec{{\rm ~erg~s^{-1}}}

\def\kms{~km~s$^{-1}$}

\def\micron{$\mu$m}
\def\850{850$\mu$m}

\def\ha{{H$\alpha$}}
\def\hb{{H$\beta$}}
\def\la{{Ly$\alpha$}}

\def\oiii{[\ion{O}{III}]}
\def\nii{[\ion{N}{II}]}

\def\civ{\ion{C}{IV}}
\def\ciii{\ion{C}{III}]}
\def\niv{\ion{N}{IV}}
\def\nv{\ion{N}{V}}
\def\mgii{\ion{Mg}{II}}

\def\fulloiii{[\ion{O}{III}]$\,\lambda$5007}

\def\fullciv{\ion{C}{IV}$\,\lambda$1549}
\def\fullciii{\ion{C}{III}]$\,\lambda$1909}

\def\Lha{L$_{\rm H\alpha}$}

\def\.25{0.25 keV\thinspace}

\def\z2{z$\,\sim\,2$}

\def\kb{K$_{\rm b}$}
\def\fwhmha{FWHM$_{\rm H\alpha}$}

\newcommand\dummytable{\refstepcounter{table}}%

\begin{document}
   \title{Sub-millimeter Detected z $\sim$2 Radio-quiet QSOs}
   \subtitle{Accurate Redshifts, Black Hole Masses, and Inflow/Outflow Velocities}

   \author{Gustavo Orellana\inst{1}
          \and
          Neil M. Nagar\inst{1}
          \and
          Kate G. Isaac\inst{2}
          \and
          Robert Priddey\inst{3} 
          \and 
          Roberto Maiolino\inst{4} 
          \and 
          Richard McMahon\inst{5} 
          \and 
          Alessandro  Marconi\inst{6,7} 
          \and 
          Ernesto Oliva\inst{7}
          }
   \offprints{G. Orellana}

   \institute{Astronomy Department, Universidad de Concepci\'on, Concepci\'on, Chile \\ 
              \email{gorellana@udec.cl,nagar@astro-udec.cl}
         \and
             ESA Astrophysics Missions Div, ESTEC/SRE-SA Keplerlaan 1, NL-2201 AZ, Noordwijk, The Netherlands \\
         \and 
             Centre for Astrophysics Research, University of Hertfordshire, College Lane, Hatfield AL10 9AB \\
         \and 
             INAF-Osservatorio Astronomico di Roma, via di Frascati 33, 00040 Monte Porzio Catone, Italy 
         \and
             Institute of Astronomy, University of Cambridge, Madingley Road, Cambridge CB3 OHA  \\
         \and 
             Dipartimento di Fisica e Astronomia, Universit\'a degli Studi di Firenze, Largo E. Fermi 2, Firenze, Italy
         \and
             INAF-Osservatorio Astrofisico di Arcetri, Largo E. Fermi 5, 50125 Firenze, Italy  
             }
   \date{Received September 22, 2010; accepted April 05, 2011}

  \abstract
   {We present near-IR spectroscopy of a sample of luminous 
    ($\rm{M_B<-27.5}$; $\rm{L_{bol}>10^{14}}$~\lsun), sub-millimeter-detected, dusty
    ($\rm{M_d}\sim10^9$~\msun), radio-quiet quasi-stellar objects (QSOs) at \z2. 
   }
   {A primary aim is to provide a more accurate QSO redshift determination in order to
    trace kinematics and inflows/outflows in these sub-mm bright QSOs.
    Additionally, the \ha\ and continuum properties allow an estimation of the
    black hole mass and accretion rate, offering insights into the 
    starburst-AGN connection in sub-mm bright QSOs. 
   }
   {We measure the redshift, width, and luminosity of the \ha\ line, and the continuum 
    luminosity near \ha. 
    Relative velocity differences between \ha\ and rest-frame UV emission lines 
    are used to study the presence and strength of outflows/inflows.
    Luminosities and line widths are used to estimate the black hole masses,
    bolometric luminosities, Eddington fractions, and accretion rates; these are 
    compared to the star-formation-rate (SFR), estimated from the sub-mm derived
    far-infrared (FIR) luminosity. 
    Finally our sub-mm-bright QSO sample is compared with other QSO samples 
    at similar redshifts. 
   }
   {The \ha\ emission line was strongly detected in all sources. Two components --
    a very broad ($\gtrsim$5000\kms) Gaussian and an intermediate-width ($\gtrsim$1500\kms) 
    Gaussian, were required to fit the \ha\ profile of all observed QSOs. 
    Narrow ($\lesssim$1000\kms) lines were not detected in the sample QSOs.
    The rest-frame UV emission lines in these sub-mm bright QSOs show larger than average 
    blue-shifted velocities, potentially tracing strong -- up to 
    3000 \kms\ -- outflows in the Broad Line Region.
    With the exception of the one QSO which shows exceptionally broad \ha\ lines, the 
    black hole masses of the QSO sample are in the range 
    log~M$_{\rm BH}$ = 9.0--9.7 and the Eddington fractions are between 0.5 and $\sim$1. 
    In black hole mass and accretion rate, this sub-mm bright QSO sample
    is indistinguishable from the Shemmer et al. (2004) optically-bright QSO sample at \z2;
    the latter is likely dominated by sub-mm dim QSOs.
    Previous authors have demonstrated a correlation, over six orders of magnitude,
    between SFR and accretion rate in active galaxies: 
    the sub-mm bright QSOs lie at the upper extremes of both 
    quantities and their SFR is an order of magnitude higher than that predicted from the
    correlation.
   }
   {
    }

   \maketitle
%
\section{Introduction}
\label{secintro}

There is now general agreement that dormant black holes, believed to be QSO `relics',
are common, perhaps ubiquitous, in nearby galactic nuclei 
\citep[e.g.,][and references therein]{sol82,fermer00,treet02,yutre02,maret04,merhei08,sha09}.
The correlation between the mass of the black hole and the structural
parameters of the galaxy spheroid  \citep[$M_{bh} \approx 6\times10^{-3}M_{sph}$, 
e.g.,][]{kor95,mag98,geb00,treet02,maret03,haring04,hop07,gulte09}
suggests an intimate connection between the existence of
a luminous QSO phase and the formation of massive, early-type galaxies.
The peak in the space density of QSOs at z$\sim$2 
\citep{schet91,walet05} suggests that these massive black holes formed at early epochs.
It seems that {\it all} sufficiently massive bulges
undergo a QSO phase with the AGN fueled by the initial collapse of
the host object at high redshift.
This picture is supported by the detection of massive elliptical
galaxies hosting all luminous, nearby QSOs \citep{bahet97,mclet99}.
Any account of bulge formation evolution which disregards the coevolution
of the central black hole -- or {\it vice versa}-- is fundamentally
incomplete \citep{kauhae00,sha09}. 
A continuing task is to unify different galaxy types with varied luminosities 
into a common evolutionary scenario or parameter space. 

Over the last few years we have pursued a highly successful program aimed at 
elucidating the star formation history of optically-bright radio-quiet QSO host galaxies
\citep{priet07,robet04,priet03a,omoet03,priet03b,isaet02,omoet01,mcmet99,omoet96a,mcmet94,isaet94}
via the detection of thermal emission from cool ($\sim$50~K) dust with 
SCUBA, MAMBO and their single-pixel/few-pixel predecessors. 
Optically-bright QSOs were targeted as likely signposts to 
massive underlying host galaxies in an era before large blind sub-mm surveys were possible.
This program concentrated on radio-quiet (to minimize the contamination of synchrotron
emission to the sub-mm fluxes) QSOs at both the highest redshift (z$\gtrsim$4) 
and at the epoch of peak QSO activity (z$\sim$2). 
Typical star formation rates (SFRs) - estimated from the conversion of sub-mm fluxes into
far-infrared (FIR) luminosities and then SFRs \citep[e.g.,][]{mcmet99} -  
of $\sim500$~\msun/yr were found in sub-mm detected QSOs, although rates upto 
$\sim2500$~\msun/yr are observed in some cases, e.g., the spectacular merging system in
the z=4.7 QSO BR1202$-$0725. Note that these SFRs are estimated under the assumption
that the sub-mm sources are not lensed.
This intense star-formation scenario is supported by the detection of the essential 
molecular material \citep[e.g.,][]{solvan05,maiet07,copet08}, and the presence of luminous 
Polycyclic aromatic hydrocarbon (PAH) emission \citep{lutet08} in several of the sub-mm detected QSOs.
The correlation between PAH emission and sub-mm emission
\citep{lutet07,lutet08} supports star formation related heating of the dust rather
than AGN heating.
The studies of optically bright, radio-quiet QSOs at z$\sim$2 \citep{priet03a,omoet03} were the 
first systematic studies of the star-formation properties of QSOs at the epoch that spans
the peak of AGN activity \citep{schet91,walet05} and 
resulted in the mm and sub-mm detection of 18 QSOs in the redshift range 1.5 $\leq$ z $\leq$ 3.0. 
The observing approach adopted in the SCUBA (850\micron) survey was to identify the very
brightest (\850\ flux $\gtrsim$7mJy)
of the sub-mm-bright QSO host galaxies with the aim of establishing a small sample
suitable for follow-up observations in the pre-ALMA era to search for the molecular
gas reservoir expected in these prodigiously star-forming galaxies.
Assuming accretion onto a black hole at the Eddington rate, with efficiency $\epsilon\sim0.1$,
the typical observed z$\sim$2 optically-bright ($B$-band absolute magnitude $\sim$-$27.5$) QSO is hosted by  
a spheroid of mass $M_{sph}\approx10^{11}$M\sun.

Previous redshifts of QSOs in the z$\sim$2 sub-mm detected sample were based
on broad (5,000-10,000\kms) high-ionization rest-frame UV
lines that are observed in the optical, e.g., \civ, \ciii, \la, and \niv.
Different broad emission lines in an ensemble of QSOs 
show systematically different redshifts, with the high-ionization lines 
(e.g., \fullciv, \fullciii) displaying mean velocity differences of $\sim$500\kms\ 
and dispersions (1$\sigma$) about this mean of 100$-$200\kms\  
\citep{tytfan92,vanet01}.
In addition it is not clear if the broad lines trace the
systemic redshift of the host galaxy (as represented by the stellar content
of the galaxy). \citet{pen77} proposed that the gas which emits narrow forbidden
lines (e.g., \oiii) in active galaxies is
likely to give a good estimate of the systemic center-of-mass
redshift. This low density gas comes from a region $\sim$1~kpc in extent
and its motion should be dominated by the stellar density field in
contrast to the broad line region gas which lies within a few parsecs of
the central black hole. 
In local active galaxies, redshifts from these
narrow ($\lesssim$1000\kms) forbidden lines have shown agreement to
$<$100\kms\ of the accepted systemic rest-frame determined from
stellar absorption features and HI 21~cm emission in host galaxies.

Broad permitted lines in QSOs, both rest-frame UV lines \citep{broet94} 
and \hb\ \citep{huet08}, can often be decomposed into a 
very broad component ($\gtrsim$5000 \kms) and an intermediate-width 
component ($\sim$1500-3000 \kms; typically 0.4 times the width of
the very broad component). 
The corresponding intermediate line region (ILR) is interpreted to be the outer
regions of the broad line region (BLR) and may be dominated by infall \citep{huet08}.
The velocity shift between the ILR emission lines and the \oiii\ 
lines is typically within a few hundreds of \kms\ \citep{huet08}. The former is 
thus a better tracer of the QSO systemic velocity in comparison to rest-frame UV lines. 

\begin{figure*}[ht]
\includegraphics[bb=0 320 612 460,width=\textwidth,clip]{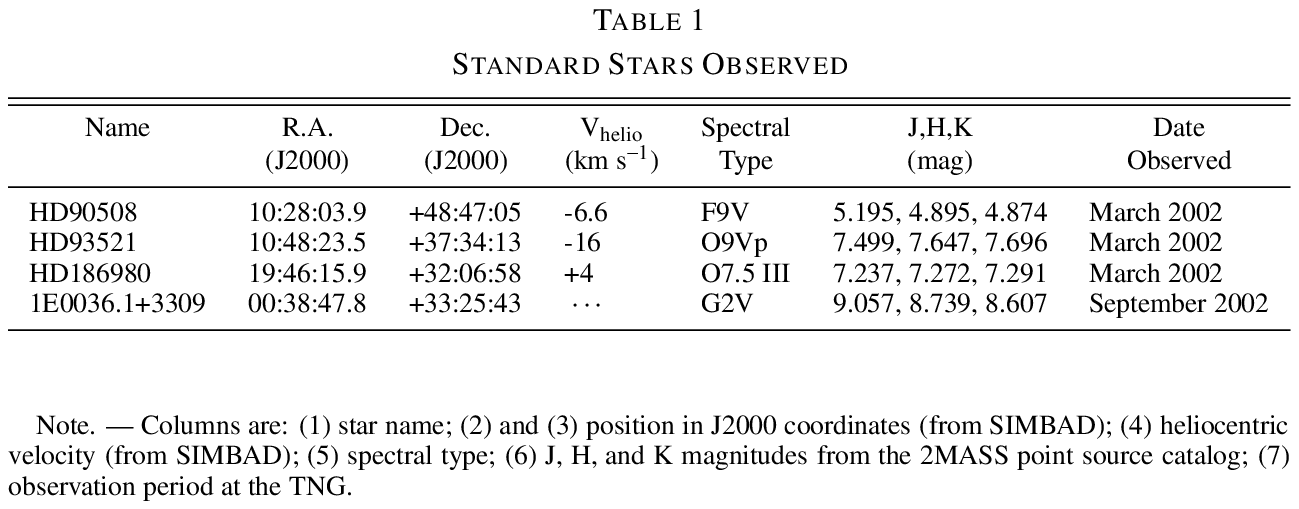}
\end{figure*}

Black hole masses in QSOs can be estimated via scaling relations derived from
reverberation mapping \citep{blamck82,vespet06} results: specifically via a single
epoch measurement of a continuum luminosity \citep[representing BLR size;][]{kaset00}
and the FWHM of a broad permitted line, e.g., \hb, \ha, \civ\
\citep[representing gas velocity in the BLR;][]{benet09}.
The most direct and best characterized single-epoch method, derived from 
reverberation mapping results in nearby (relatively low luminosity) Seyfert galaxies, 
is via the pair L$_{\rm 5100}$ and FWHM$_{\rm H\beta}$. The single-epoch
black hole mass estimation based on this pair is expected to be accurate to a factor
$\sim$2, \citep[e.g.,][]{benet09}.
The extrapolation of this black hole mass estimator to high luminosity
QSOs has been questioned on two main fronts: 
(a) is the relationship
  between continuum luminosity and BLR size 
  \citep[the $R_{\rm BLR} - L$ relationship;][]{kaset05} unchanged at higher luminosities?
  Initial results with the \civ\ line \citep{kaset07} imply that it is relatively unchanged
  though a firm result using the \hb\ line is awaited \citep[][]{botet10}.
  The results in this work are relatively unaffected by small changes in the relationship
  -- currently tested to continuum luminosities up to Log~($\lambda$ L$_\lambda$) 
  = 46 \citep{benet09} -- as the QSOs discussed here have continuum luminosities
  of Log~($\lambda$ L$_\lambda$) $\simeq$ 46.3--46.8; 
(b) should the FWHM of the \hb\ line be estimated only from the Very Broad Line Region (VBLR) profile or
   from the combined VBLR and ILR profile? It is not uncommon that reverberation-mapped Seyferts show
   complex broad line profiles which cannot be fit with a single broad Gaussian. Further, complex gas motions
   including inflows and outflows are seen in these Seyferts \citep[e.g.,][]{denet09}. Even in these 
   complex cases it appears that
   using a simplistic FWHM for the full \hb\ profile (after subtraction of only
   the NLR emission) does not appreciably change the scaling relations used to 
   estimate the black hole mass \citep{denet10}. On the other hand, there is
   some evidence that only a few local Seyferts (2 of 10 studied) show a true ILR 
   component; for these, the inclusion of the ILR component in the FWHM measurement
   leads to erroneous black hole estimations \citep{zhuet09,zhuzha10}.

An initial motivation of this work was the derivation of an accurate QSO redshift
in order to obtain molecular gas detections on instruments with bandwidth 1-2~GHz. This
required redshifts accurate to, e.g., 1\% at the 115~GHz CO J:1-0 line. 
The recent availability of extremely wide-band receivers, e.g., Z-SPEC, 
ZEUS, GBT-Spec, has now relaxed this requirement for a very accurate redshift.
Rest-frame optical spectroscopy of these unique sub-mm-bright
QSOs also permits constraints on their black hole masses and accretion rates, thus
allowing comparisons between galaxy and black hole growth.
Further, an accurate systemic velocity allows better constraints on the kinematics
of the nuclear gas.

In this work we present near-IR spectroscopy probing the 
\ha\ line and nearby continuum in ten of the 18 QSOs
in a z$\sim$2 sub-mm detected QSO sample.
The sample selection, observations, and data processing are
summarized in Sect.~\ref{secsample}, results are presented in
Sect.~\ref{secresults}, and a brief discussion and concluding
remarks are presented in Sect.~\ref{secdiscussion}.
Luminosity distances of the QSOs were calculated using a 
Hubble constant $H_0$= 72 \H0, $k = 0$, $\Omega_{\rm m} = 0.3$, 
$\Omega_{\Lambda}= 0.7$, and q0=$-0.5$.
  
\section{Sample, Observations, and Data Processing}
\label{secsample}

The sample is drawn from \citet{priet03a} and \citet{omoet03}.
\citet{priet03a} surveyed 57 optically luminous
(M$_{\rm B} < -27.5$), radio-quiet QSOs 
at z$\sim$2 from the Large Bright Quasar Survey \citep[LBQS;][]{hewet95} and 
the Hamburg Bright Quasar Survey \citep[HBQS;][]{haget99,enget98,reiet95} at  
\850\ with JCMT/SCUBA. 
Here radio quiet was defined by a non-detection in the NRAO VLA Sky Survey (NVSS), 
i.e. S$_{\rm 1.4GHz} <$ 1.5~mJy\footnote{The assumption of a spectral index
$\alpha~=-0.5$ or steeper would imply S$_{\rm 850GHz} <$ 0.1~mJy.}.
\citet{priet03a} detected 9 QSOs at a significance of 3$\sigma$ or greater; these
have  850\micron\ fluxes in the range 7--17~mJy. The remaining QSOs have 
3$\sigma$ flux density limits in the range 6--10~mJy.
\citet{omoet03} surveyed 35 optically luminous (M$_{\rm B} < -27.0$)
radio-quiet QSOs with $1.8<$ z $<2.8$ 
at 1.2~mm with IRAM/MAMBO and detected 9 QSOs with 1.2~mm fluxes of 3.2--10.7~mJy; the 
remaining QSOs have 3$\sigma$ flux density limits in the range 1.8--4~mJy. The
\citet{omoet03} sample was drawn from the AGN catalog of
\citet{verver00} and included a variety of QSO surveys, among them
the HBQS, the LBQS,
the Second Byurakan Survey \citep[SBS;][]{steet01} and the Hewitt-Burbidge QSO
compilation \citep[HB89;][]{hewbur89}. 
Of the 83 unique QSOs surveyed by Priddey et al. and Omont et al., 
nine QSOs were observed by both. Of these nine, 
eight were not detected in both studies while HS~B1049+4033 was detected by
Omont et al. but not detected by Priddey et al. 
Thus, combining the samples of Priddey et al. and Omont et al.
resulted in 83 observed QSOs of which 18 were detected at \850\ or 1.2~mm.
Given the relatively high upper-limits of the QSOs not detected at \850\ and/or 1.2~mm
(in many cases the 3$\sigma$ flux upper-limits are higher than the fluxes of the weaker
sub-mm detected QSOs), we refrain from using the term sub-mm bright, and 
instead refer to the 18 QSOs as sub-mm detected QSOs, and the remaining 
as sub-mm non-detected QSOs. 
We have obtained near-IR spectra of ten of these 18 sub-mm detected
QSOs, listed in Table~\ref{figtab2}, which therefore form the  sub-mm detected QSO
sample discussed in this work. 

Observations were conducted on 2002 March 28, and 2002 September 20 at 
the 3.6 meter Telescopio Nazionale Galileo (TNG) at La Palma,
using the Near-IR Camera and Spectrograph (NICS) instrument
with a detector based on a HgCdTe Hawaii 1024x1024 array
\citep{bafet01}.
On 2002 March 28, we used TNG/NICS with the \kb\ grism and a slit-width of
0{\farcs}7 or 1{\arcsec}. This yielded the 
wavelength range 1.91\micron\ to 2.38\micron\ at a dispersion of 4.5\AA\ per pixel
i.e. R$\sim$1250 for a 1{\arcsec} slit. 
On 2002 September 20, we used TNG/NICS with the HK grism
and a slit of 1{\arcsec} width. This yielded the wavelength range 
1.37\micron\ to 2.49\micron\ at a dispersion of 11\AA\ per pixel. 
Standard calibration frames were taken at the beginning and end of the observing night.
The QSOs were observed using typical integration times of 120~sec
in an ``ABBA'' nodding pattern. The distance between ``A'' and ``B'' positions
was roughly 15{\arcsec}, and was varied slightly on each nod. Total integrations
were typically 30--40~min on-source per QSO.
At least two telluric `standard star' observations were obtained on each night 
(Table~\ref{figtab1}).

All spectra were first ``straightened' using 54 strong
night sky lines. Xe and Ar lamp exposures, taken at the beginning
and end of the night, were used for initial wavelength calibration,
with  individual lines identified using the line-lists (in vacuum 
wavelengths) available from the ESO/ISAAC 
website\footnote{http://www.eso.org/instruments/isaac/tools}.
The Ar spectrum is richer in lines in the case of the \kb\ grism: 
of these we used 11 reliable lines. 
The Xe spectrum contributed three additional lines to the \kb\ grism observations. 
These 14 lines ranged between 19823\AA\ and 23200\AA, which
nicely covered the full wavelength range of our setup:
$\sim$19157\AA\ to 23800\AA. Two QSOs - [HB89]0933+733 and SBS~B1408+567 -
have \ha\ lines which fell at the red-most end of the CCD: 
in these cases we checked that using additional weaker Ar and Xe lines in the red-most
region gave a similar wavelength calibration solution. 
The final wavelength solution from the lamps had an r.m.s. about 0.3\AA, with
almost all points well within 0.5\AA\ of the fit. 
In the case of the HK grism, we used only the Ar lamp, which yielded
34 usable lines, between 13179\AA\ and 23973\AA. The final fit
had an r.m.s. of 4\AA;  at the blue end
the points were all within $\pm 5$\AA\ while at the red end
the points were all within $\pm 10$\AA.
The wavelength calibration was checked both with the calibration star
spectra and with night sky lines:  in all cases the observed lines
were within 1\AA\ (March 2002) or 3\AA\ (September 2002) of their
expected values.
In the case of data taken in March 2002, the expected wavelength of the
night sky lines was shifted by a consistent amount (within 3\AA) in
observations made towards the end of the night. 
The latter shift was probably due to a slight change in tilt
of the grism with telescope movement, and we corrected the wavelength solution 
of the target objects with an equivalent zero-point shift. The accuracy of
the wavelength calibration was therefore within 2\AA\ or $\sim$30\kms\
at the typical observed wavelengths of the \ha\ line.

Finally, QSO spectra were continuum subtracted assuming a linear baseline
and the \textit{Specfit} \citep{kri94} package within IRAF
was used to obtain Gaussian and Lorentzian fits to the \ha\ line profile.
The very broad profiles of the lines results in an uncertainty of about 5--10\AA\ 
in the fitted central wavelengths (10--15\AA\ for the
September 2002 data due to its lower spectral resolution). 
This is the dominant source of measurement error in the redshift determinations. 

We note that NICS data was taken soon after instrument commissioning,
at a time when telescope and instrument pointing was still
being fine tuned. The telescope nodding along the slit was therefore slightly
inaccurate, and the counts on the standard stars and QSOs varied, in
the worst case, by up to 20\% between exposures. Flux calibration was
performed by comparing the best spectra of the standard stars to the
expected spectra as estimated from black body fits to the V- to K-band magnitudes
of the standard stars. 
The flux calibrated QSO spectra were then used to estimate the J, H, and K-band
magnitudes of the QSOs. We found 
that the spectral fluxes had to be increased by $\sim$30\% in order
to be consistent with the observed 2MASS magnitudes of the QSOs. This correction, likely 
the result of slit losses, was thus applied to all QSO spectra.

We use the primary QSO samples of \citet{sheet04}, \citet{netet07} and \citet{aleet08}
to compare and contrast the properties of our QSO sample.
\citet{sheet04} present the results of new rest-frame
optical spectra of a sample of 29 high-redshift (typically
z$\sim$2.3) luminous (L $\geq 10^{46} \ergsec$) QSOs and
\citet{netet07} present the results of new rest-frame optical spectroscopy of 15
luminous QSOs at z$\sim$2.3--3.4. \citet{aleet08} present 
a sample of z$\sim$2 sub-millimeter detected
galaxies (SMGs) which exhibit broad \ha\ or broad \hb\ emission. 
All three works use the results of the rest-frame optical
spectroscopy to derive estimates of the black hole mass and related
quantities  which are directly comparable to our results. 
In all cases
we convert listed fluxes into luminosities using the cosmological
parameters of this work.
We also use the results of \citet{copet08} who present rest-frame
optical spectroscopy of a sample of sub-mm detected QSOs, four of which are
also in our sample.

\section{Results}
\label{secresults}

\subsection{\ha\ Profile Fits}
\label{secprofile}

\begin{figure*}[ht]
\includegraphics[bb=100 90 640 420,width=1.0\textwidth,clip]{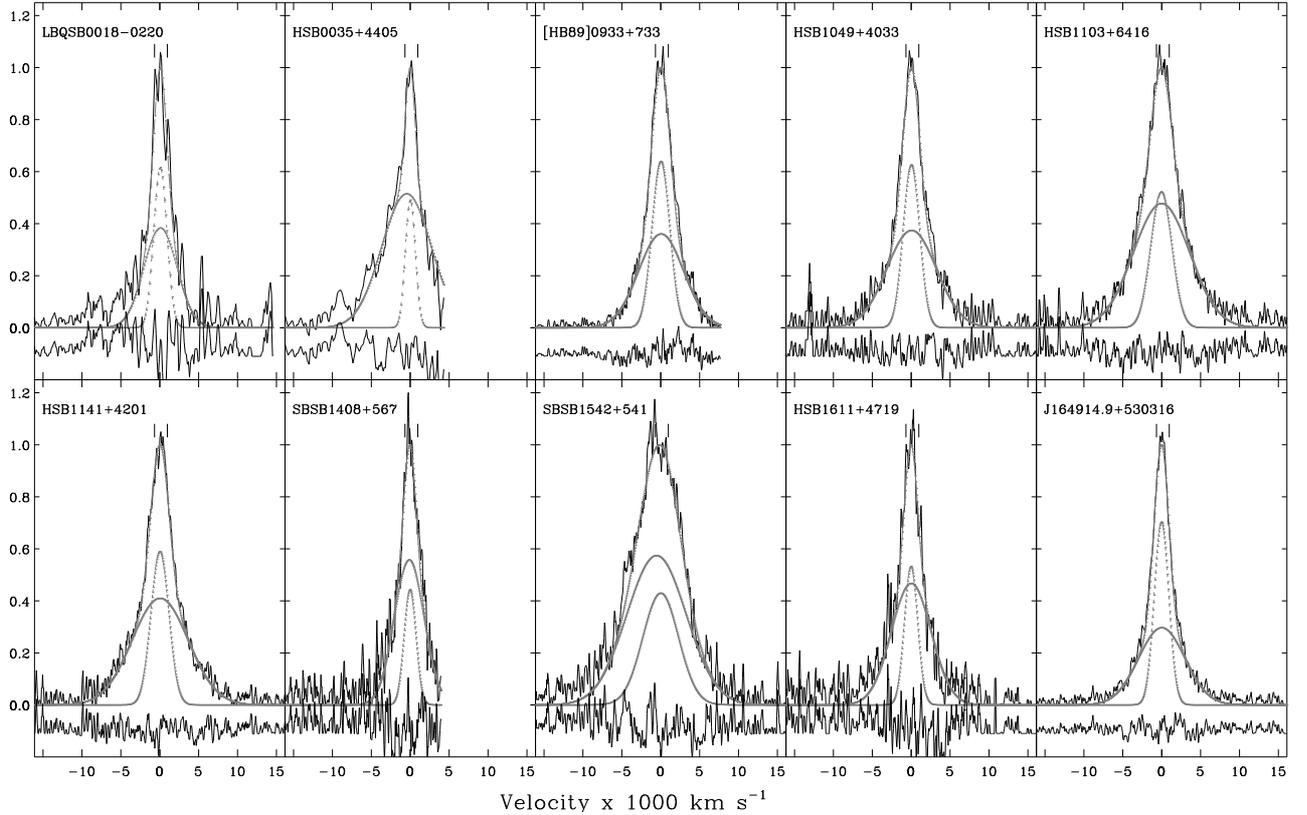}
\caption{Near-IR spectra of the sample QSOs centered on the 
\ha\ emission line, along with our double-Gaussian fits to the line profile. 
The two Gaussians represent an intermediate-width component (corresponding to
the ILR), and a very broad component (corresponding to the VBLR; see text).
For each QSO the upper panel shows the normalized QSO spectrum (black) 
along with the double-Gaussian fit (grey); 
the lower panel shows the residual spectrum after subtraction of the double-Gaussian fit.
The zero velocity is based on the central wavelength of the intermediate-width component 
of the double-Gaussian fit (see Table~\ref{figtab2}).
The ticks below each QSO name indicate the expected locations of the \nii\ doublet.}
\label{figfit}
\end{figure*}

Spectra covering the region near the \ha\ line are shown in Fig.~\ref{figfit} 
for all observed QSOs.  None of the sample QSOs have \ha\ profiles which 
can be satisfactorily fit with a single Gaussian.
Two components, a very broad (4500--9000~\kms) Gaussian and an intermediate-width
(typically 1500-3200~\kms) Gaussian, are instead required. Narrow lines 
($\lesssim$1000\kms) are not clearly detected in any of the sample QSOs, 
though there is a very tentative detection of narrow \ha\ + \nii\ lines
in LBQS B0018-0220 (Fig.~\ref{figfit}).
Both the very broad and the intermediate-width Gaussian components are centered on 
more or less the same wavelength (differences of $\leq$10~\AA\ or $\leq$120\kms),
for all QSOs except  SBS~B1542+541, which has exceptionally wide
ILR and VBLR \ha\ components and an obvious blue shoulder,
and HS~B0035+4405, for which a reliable double-Gaussian fit was
difficult given that the red wing of the \ha\ profile extended beyond
the observed range. In the two latter cases, the difference between the
fitted central wavelengths of the very broad and intermediate-width Gaussian components, 
40\AA, represents a velocity difference of about 500\kms. 
On the other hand, a single Lorentzian profile provides a satisfactory fit 
to the bulk of the \ha\ line profile in most QSOs, with the main drawback of 
potentially overestimating the line profile in the outer wings. 
This single Lorentzian fit to the \ha\ profile has a central wavelength 
similar to, or the same as, that of the intermediate-width Gaussian component in 
the double-Gaussian fit.

We calculate the systemic redshift of the QSO from the central wavelength of the 
intermediate-width (ILR) component of the double-Gaussian fit to \ha. 
The total flux of the \ha\ line is taken as the summed flux of the 
double-Gaussian fit. Given our non-detection 
of any narrow (NLR) component, the measured \ha\ fluxes and widths are dominated by the BLR 
(VBLR plus ILR).
The estimation of black hole masses for the sample QSOs requires the measurement
of the full-width at half maximum (FWHM) of the broad component of the \ha\ emission
profile, i.e. the FWHM of the \ha\ profile after subtraction of any narrow 
($\lesssim$1000 \kms) component. Given that a single Gaussian does not provide
a good fit to the line profile and that the narrow component is not detected
in any QSO, we calculate the FWHM of the \ha\ line profile 
from the summed double-Gaussian fit, and refer to this quantity
as \fwhmha. This \fwhmha\ is within 50--100\kms\ of the FWHM of the single Lorentzian fit
for all QSOs except SBS~B1542+541 for which the difference is 620\kms\ 
(Table~\ref{figtab2}).

The double-Gaussian fits to the \ha\ lines are shown in Fig.~\ref{figfit} for
the sample QSOs, and the corresponding line centers, FWHMs and fluxes 
of each Gaussian component are listed in Table~\ref{figtab2}. 
Values of the FWHM of the single-Lorentzian fit and of \fwhmha\ are also 
listed in the same table. 

\subsection{Systemic Redshifts and Velocity Offsets} 
\label{secredshift}

In this section we first compare our redshift estimations 
to previously obtained redshifts based on rest-frame-optical spectroscopy.
We then analyze differences between redshifts derived
from \ha\ and from rest-frame UV spectroscopy for the sample QSOs. 
We further study in detail the velocity offsets between individual emission lines for
the subset of the sample QSOs which have high quality rest-frame UV spectra.
Finally, we contrast the sub-mm detected and sub-mm non-detected QSOs 
in the parent samples of \citet{priet03a} and \citet{omoet03} with respect to
velocity offsets seen in different emission lines.

\begin{figure*}[ht]
\begin{center}
\includegraphics[bb=42 82 1070 570,width=\textwidth,clip]{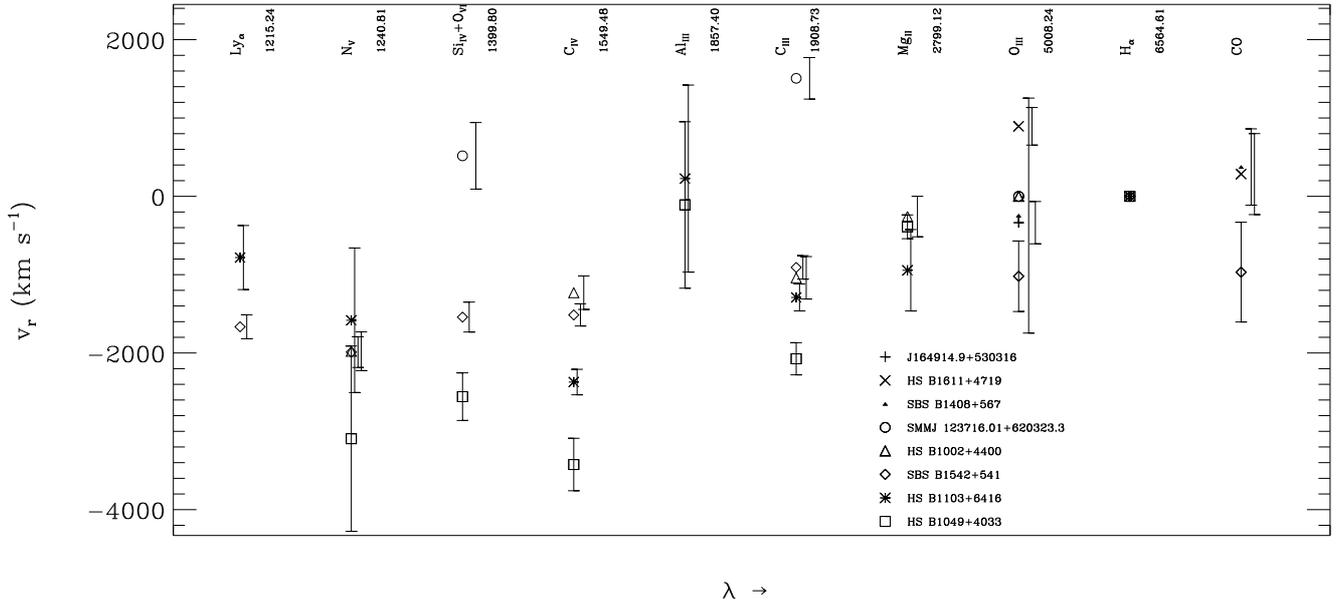}
\end{center}
\caption{Velocity offsets of rest-frame UV, rest-frame optical, and CO 
emission lines for sub-mm detected QSOs. 
The QSOs plotted are present in at least two of the following: 
our sample, the \citet{copet08} sample, and the SDSS spectroscopic database. 
The lines are ordered left to right in increasing rest wavelength, individual lines
are labeled, and different symbols, as indicated on the figure, are used for each QSO.
Only lines with reliable fits to their line centers are included (see text).
The QSO systemic velocity is assumed to be that of the intermediate-width (ILR) component of 
the \ha\ line if available, or else that of the \oiii\ line from \citet{copet08}.
Error bars have been displaced slightly along the $x$ axis for better visibility.
The QSO SBS~B1542+541 is in our sample, \citet{copet08}, and has an SDSS spectrum. 
Two QSOs, HS~B1049+4033 and HS~B1103+6416, are both in our sample and have SDSS spectra.
Two QSOs, HS~B1002+4400 and SMM~J123716.01+620323.3, are both in \citet{copet08} and 
have SDSS spectra.
Three additional QSOs, SBS~B1408+567, HS~B1611+4719, and J164914.9+530316, are both 
in our sample and that of \citet{copet08}.
}
\label{figzsdss}
\end{figure*}

Four of our sample QSOs are also in the sample of \citet{copet08}, 
who present results of near-IR spectroscopic observations which include
the \fulloiii, \hb\ and \ha\ emission lines, and observations  of the J:3-2 or J:2-1
transition of CO. 
\citet{copet08} derived QSO systemic redshifts via fits to the 
\hb\ and \oiii\ lines. A comparison between
our \ha\ determined redshifts and those of \citet{copet08} shows
an agreement of within $\sim$300\kms\ in two QSOs, but larger differences
in HS~B1611+4719 and SBS~B1542+541 (Fig.~\ref{figzsdss}). 
The latter is the QSO in which both ILR and VBLR
\ha\ components have exceptionally large widths and in which a notable
\ha\ blue shoulder is seen. Our \ha\ derived redshift is therefore
probably not as accurate as the \oiii\ and CO derived redshifts of \citep{copet08}.
In the case of HS~B1611+4719, our relatively reliable fit to the \ha\ line
results in a \ha\ redshift which is offset
by 900\kms\ from the \hb\ and \oiii\ derived redshift, but offset by only 
280\kms\ from the CO derived redshift \citep{copet08}. 

\begin{figure}[ht]
\includegraphics[bb=72 72 288 288,width=0.40\textwidth,clip]{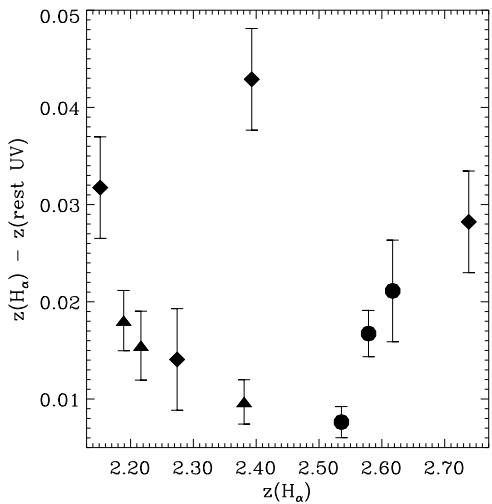}
\caption{Comparison of redshifts derived from the \ha\ line 
  and those derived from rest-frame UV spectroscopy (SDSS as triangles and 
  HBQS as diamonds). The median difference is $\Delta$z = 0.018. 
  Error bars in $y$ are indicative for non-SDSS QSOs (see text);
  error bars in $x$ are roughly the size of the symbols.
}
\label{figzhbqs}
\end{figure}

Our \ha\ derived redshifts are systematically higher 
(Fig.~\ref{figzhbqs}) than those derived from optical 
spectroscopy (i.e. based on rest-frame UV lines). 
The rest-frame UV redshifts originate from very heterogenous 
observations with diverse resolutions and spectral ranges:
these are taken from, in order of preference:
SDSS,  \cite{hewet95}, \citet{hewbur89}, \citet{steet01}, and 
the Hamburg Bright Quasar Survey \citep{haget99,enget98}. 
Of these, only the SDSS redshifts have quoted uncertainties; the other 
sources do not explicitly specify uncertainties. 
In the latter cases we use indicative uncertainties estimated from the spectral
resolution and/or the number of significant digits quoted.
The median difference is $\Delta$z = 0.018 and the mean difference is $<\Delta$z$>=0.021$, 
corresponding to $\Delta$V $\sim6300$\kms\ at wavelengths near \ha.
The largest redshift differences are typically
for z$_{\rm {rest UV}}$ from the Hamburg Bright Quasar Survey which has a low resolution 
(typically  15\AA), limited (and relatively blue) spectral coverage, and few significant
figures in the reported redshift (with no errors specified), while the 
smallest differences are typically for z$_{\rm {rest UV}}$ from the SDSS. 
Thus the redshift differences larger than $\Delta$z $\sim$ 0.02 are 
likely the result of the low spectral resolution, the limited set of lines used,
and the low precision, of the redshifts derived from rest-frame UV lines. 
Limiting ourselves to the QSOs with accurate rest-frame UV redshifts
the typical rest-frame optical to rest-frame UV redshift
difference is $\sim$0.013, or $\sim$4000\kms. 

In order to better determine the velocity offsets of individual
rest-frame UV lines we limited ourselves to the sample QSOs for which high quality
rest-frame UV spectra exist. 
To this end, we searched the Sloan Digital Sky Survey (SDSS; Data Release 7; DR7) 
archive for spectra of all QSOs in the samples of \citet{priet03a} and \citet{omoet03}
and also those of \citet{copet08}.
We use the DR7 spectra and their associated ``SpecLine'' tables which tabulate
the SDSS automated fits to all detected emission and absorption lines in the spectra.
We selected emission lines from the SpecLine tables using the following criteria:
(a)~line width ($\sigma$) between 20\AA\ and 200\AA. The lower limit avoids spurious line
detections and the upper limit avoids lines which are suspect and/or cannot yield 
sufficiently accurate line centers; 
(b)~error in $\lambda_{\rm obs}$ less than the line width ($\sigma$);
(c)~error in $\lambda_{\rm obs}$ less than 0.5\% of $\lambda_{\rm obs}$.
Since the SDSS line identification and fitting are automatic, there are some 
errors in the database of line parameters. 
We therefore continuum subtracted the SDSS spectra using a polynomial baseline fit 
and revised all lines selected above, manually re-fitting the line centers and widths 
if necessary. This led to a change in the central wavelengths and widths of some lines 
and the elimination of other lines. 
The re-fitted values of $\lambda_{\rm obs}$ and their 1$\sigma$ errors
were used to derive relative velocities between emission lines.
When a \ha\ or \oiii\ derived redshift was available, this was used
as the systemic redshift; else only relative velocities between rest-frame UV lines
were analyzed.

\begin{figure*}[ht]
\begin{center}
\includegraphics[bb=92 72 422 268,width=0.33\textwidth,clip]{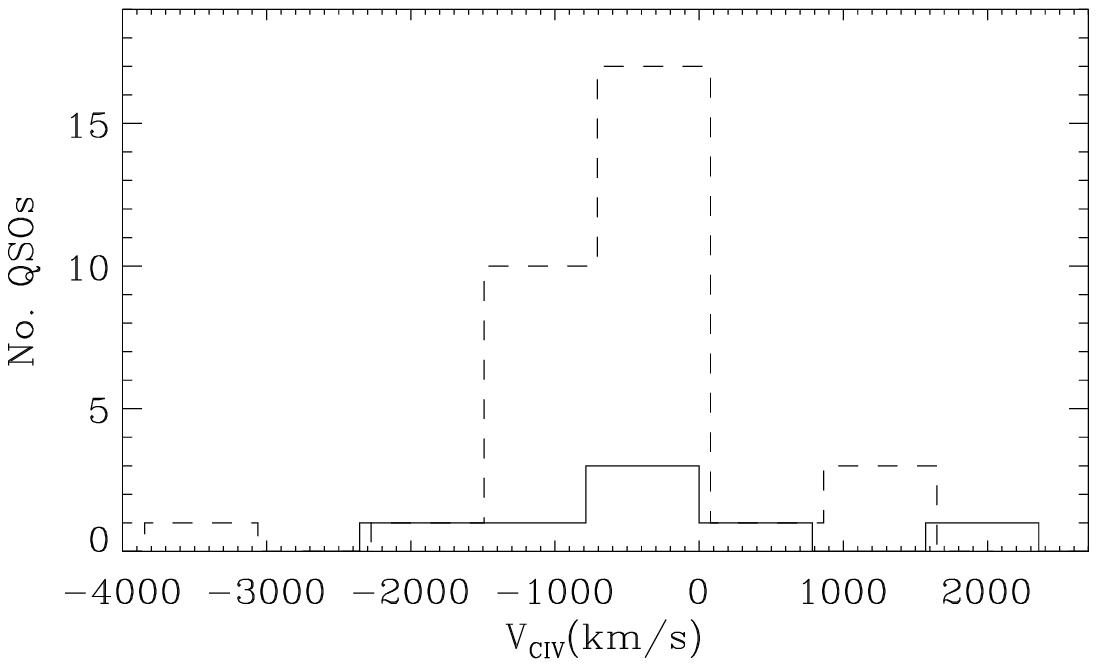}
\includegraphics[bb=92 72 422 268,width=0.33\textwidth,clip]{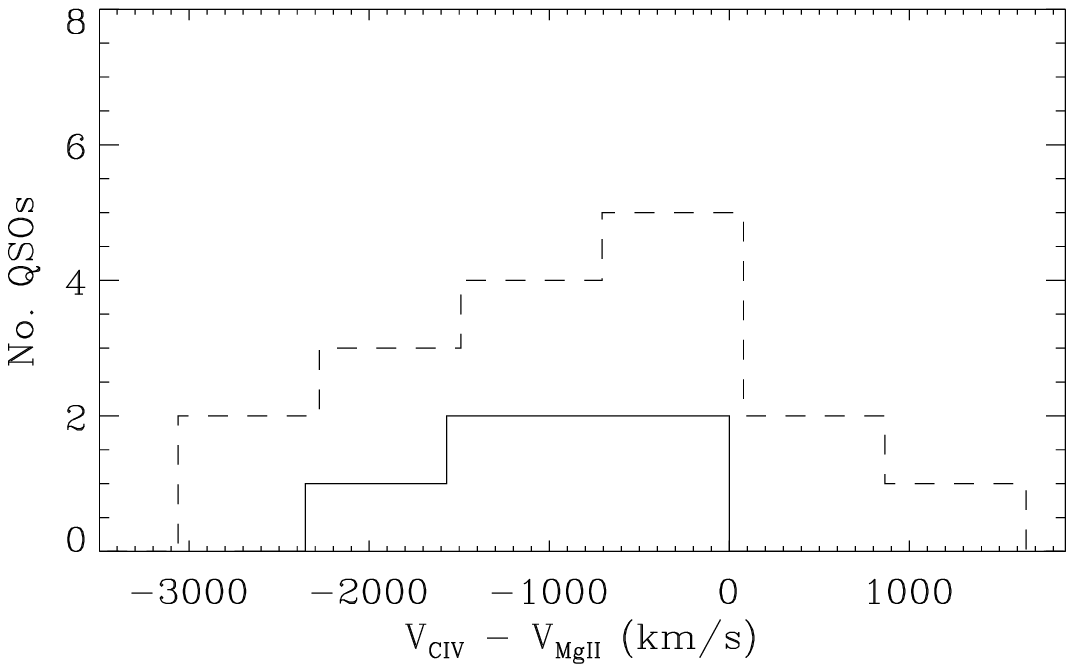}
\includegraphics[bb=92 72 422 268,width=0.33\textwidth,clip]{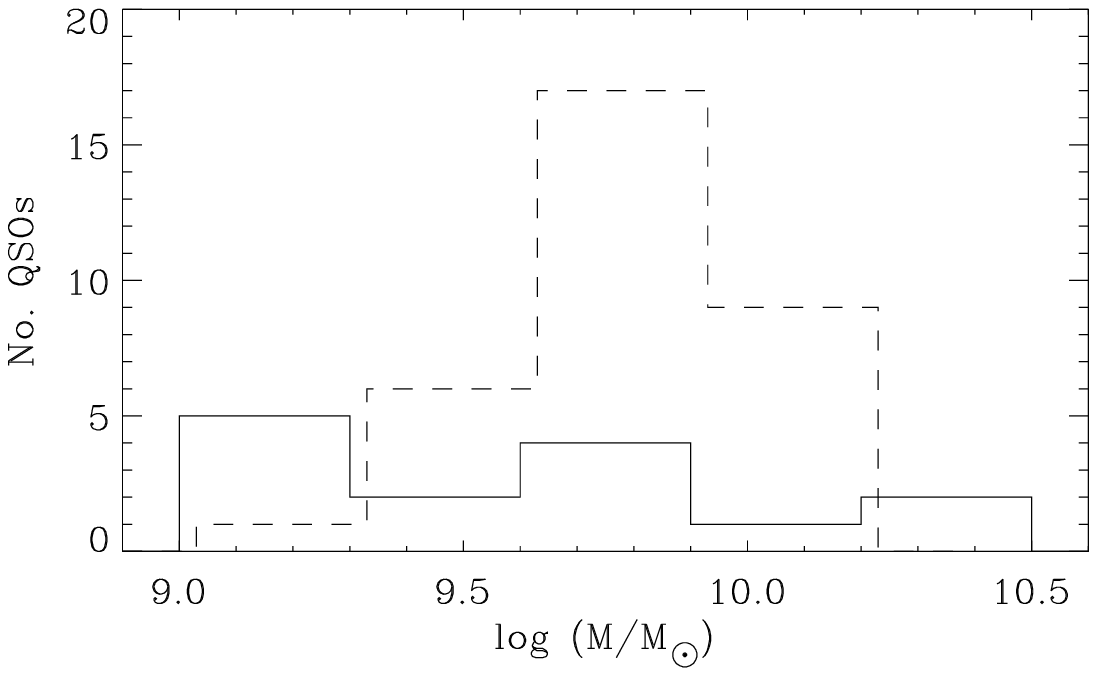}
\end{center}
\caption{A comparison of the distributions of 
\emph{Left:} \civ\ velocity relative to the SDSS-derived systemic velocity of the QSO,
\emph{Middle:} \civ\ velocity relative to the \mgii\ velocity, and 
\emph{Right:} black hole mass, 
contrasting the sub-mm detected QSOs (solid lines) and sub-mm non-detected QSOs (dashed lines)
from the samples of \citet{priet03a} and \citet{omoet03}. 
The histograms have been shifted slightly in the $x$-axis for legibility. Values used are 
from the compilation of \citet{sheet10} except for the inclusion of the 10 \ha-based black 
hole masses presented in this work. In contrast to \citet{sheet10} we use positive velocities
to indicate redshift.
}
\label{fighisto}
\end{figure*}

A comparison of the velocity offsets between emission lines for the 
QSOs in our sample and those of \citet{copet08} are shown in Fig.~\ref{figzsdss}. 
Almost all rest-frame UV emission lines show large blue-shifted velocities;
several factors higher than typically seen in QSOs \citep{tytfan92,vanet01},
with the most extreme blue-shifted velocities seen in HS~B1049+4033. We note that
HS~B1049+4033 is not extreme in its sub-mm properties: it was detected at 
1.2~mm \citep[3.2~mJy;][]{omoet03} but not detected at 
850\micron\ \citep[3.9$\pm$3.2~mJy;][]{priet03a}.
\citet{vanet01} present median velocities for emission lines relative
to \fulloiii\ for over 2000 QSOs from the SDSS over the redshift range 
$0.04~<~z~<~4.8$. Relative to the \oiii\ line, they find median velocity 
differences of 27~\kms\ for \ha, $-$70~\kms\ for \nv, $-$563~\kms\ for \civ, 
$-$224~\kms\ for \ciii, and 161~\kms\ for \mgii. 
Relative to \ha\ (or in the absence of \ha, relative to \oiii) derived
redshifts, the sub-mm detected QSOs in Fig.~\ref{figzsdss}
have \civ\ velocities of $-1230$~\kms\ to $-3420$~\kms, and 
\ciii\ velocities between $-$900~\kms\ to $-$2070~\kms\ except for
SMMJ~123716.01+620323.3 which has a \ciii\ velocity of
1500~\kms\ relative to the \oiii\ systemic velocity 
published by \cite{copet08}. 
Note that the sample QSOs show small blue-shifted velocities for \mgii\ and \la\ instead of the
small median redshifted velocities seen in the QSO sample of 
\citet{vanet01}; the \la\ velocity offsets seen in Fig.~\ref{figzsdss} 
are not reliable as the \la\ line center is difficult to ascertain given its
blue-ward absorption and the contamination from the \nv\ line. 

In the case of the parent samples of 
sub-mm detected QSOs vs. sub-mm non-detected QSOs \citep{priet03a,omoet03}, 
40 of the 89 QSOs have SDSS spectra and also appear in the compilation of 
\citet{sheet10}. 
In this larger sample, we do not have reliable QSO systemic velocities via the \ha\ or 
\oiii\ line for all QSOs and therefore only tested for relative velocity 
offsets between rest-frame UV lines. No significant difference between
the sub-mm detected and sub-mm non-detected sub-samples was found.
We also used the velocity offsets of the \civ\ and \mgii\ emission lines listed in the
\citet{sheet10} compilation of QSO properties (these are
the only two rest-frame UV emission lines for which velocity offsets from the 
SDSS-derived systemic QSO velocity are listed in the compilation). 
Note that \citet{sheet10} used negative velocities
to indicate redshift, while we use positive velocities to indicate redshift.
The velocity offsets of the \civ\ lines are not significantly different for
the sub-mm detected and non-detected subsamples (left panel of Fig.~\ref{fighisto}). 
Velocity offsets for the \civ\ and \mgii\ lines were calculated relative to the SDSS-derived 
redshift of the QSO. This SDSS-derived redshift -- based on a cross correlation of
narrow emission and absorption lines -- does not necessarily trace the systemic
redshift of the QSO as it is likely biased by the large velocity offsets
in rest-frame UV emission-lines as discussed previously.
The \mgii\ line typically shows velocity offsets which are smaller than
those of the \civ\ line, e.g., \citet{vanet01} and Fig.~\ref{figzsdss}.
We thus also compare the velocities of the \civ\ lines relative
to those of the \mgii\ lines, which potentially better represent the  
velocity offset of the \civ\ lines from the true systemic QSO redshifts
(middle panel of Fig.~\ref{fighisto}). While both subsamples show large blue-shifted
velocities, there is no clear difference between the two subsamples.  

\subsection{Black Hole Masses and Related Properties} 
\label{secbhmass}

\begin{figure*}[ht]
\includegraphics[bb=85 60 475 490,width=0.33\textwidth,clip]{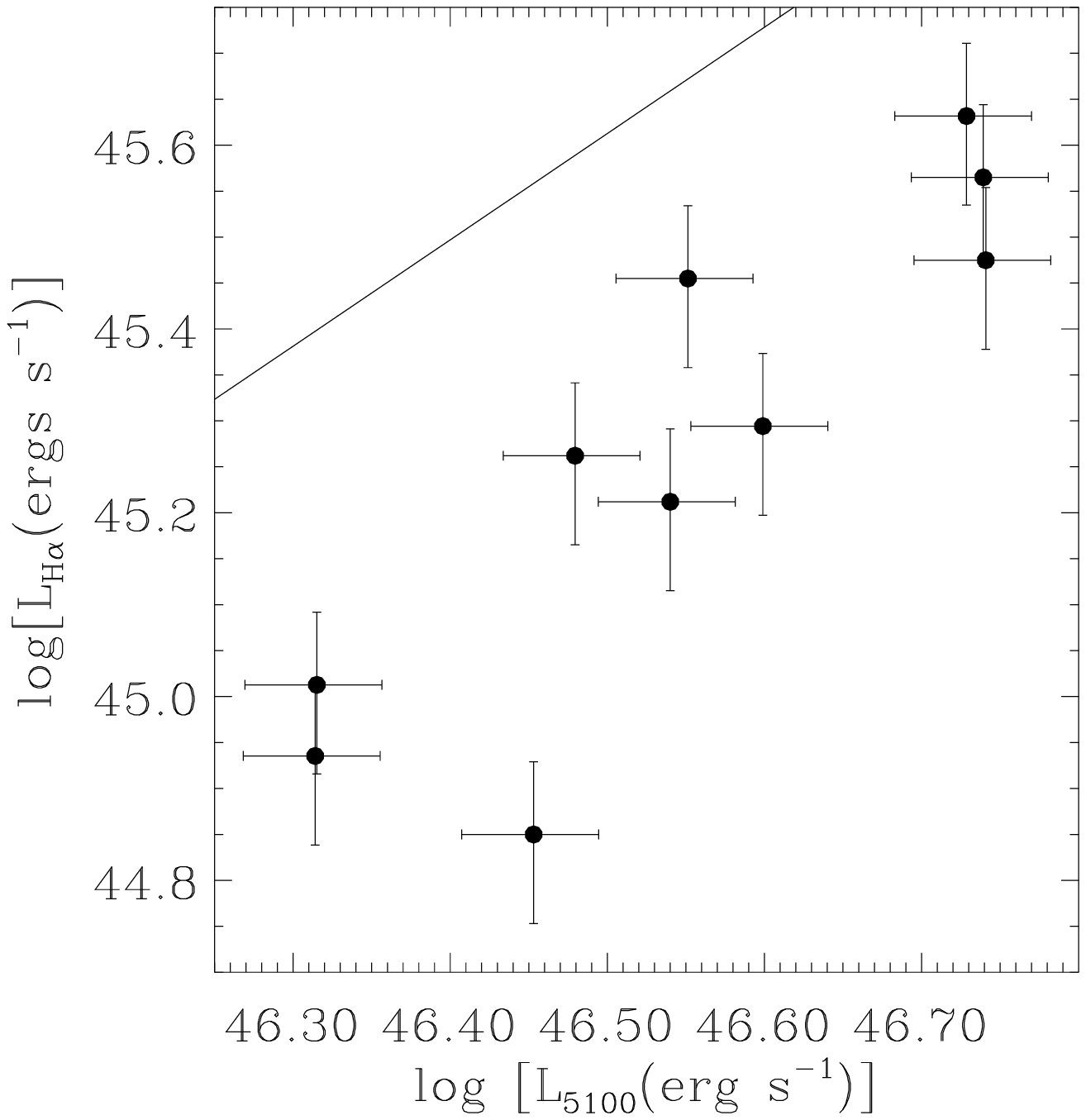}
\includegraphics[bb=85 60 475 490,width=0.33\textwidth,clip]{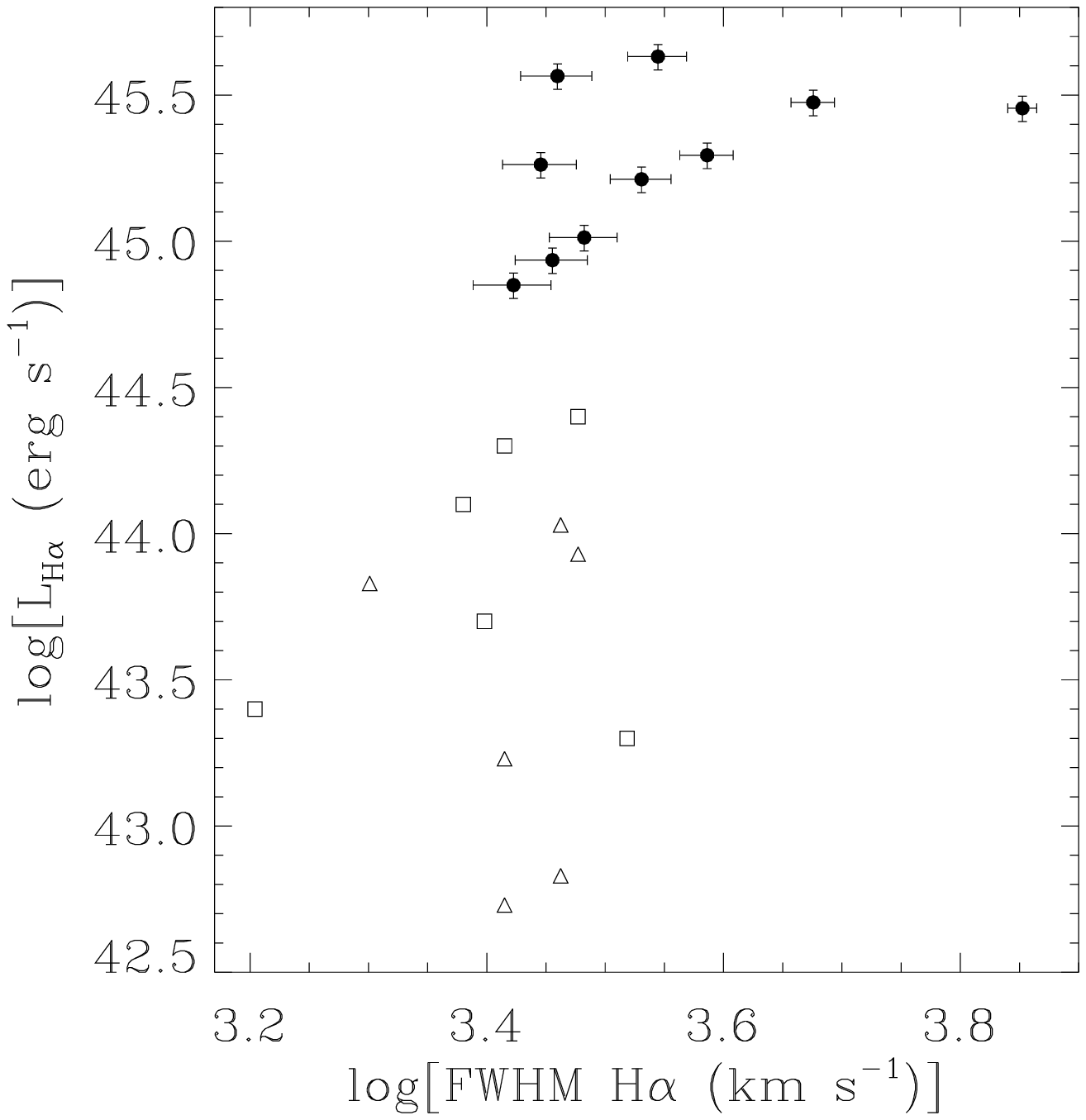}
\includegraphics[bb=85 60 475 490,width=0.33\textwidth,clip]{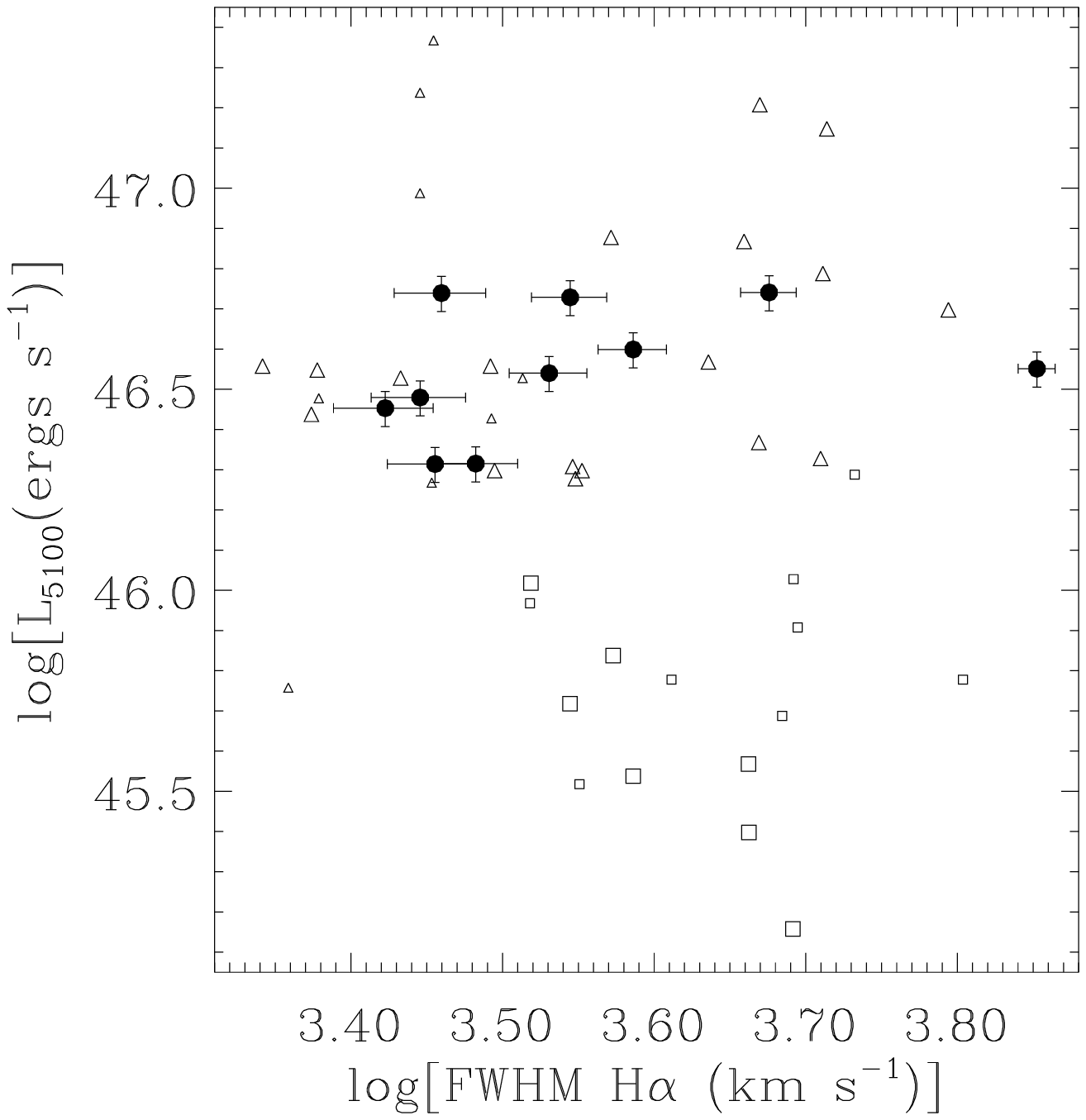}
\caption{The relationship between continuum luminosity ($\lambda$ L$_\lambda$
at 5100\AA; see text), and the width and luminosity of the broad component (ILR+VBLR)
of the \ha\ line. QSOs in our sample are plotted with filled circles and their corresponding
1$\sigma$ error bars. 
The line in the left panel corresponds to the relationship between these quantities found by
\citet{greho05} in a sample of SDSS galaxies (see text). In the middle panel
data from \citet{aleet08} on the sub-mm 
detected (SMG) galaxies in the Chandra Deep Field North are shown as triangles 
(obscured ULIRGs with broad Pa$\alpha$) and squares (SMGs with broad \ha\ or broad \hb). 
The right panel includes the QSO samples of \citet[][triangles]{sheet04} and 
\citet[][squares]{netet07}; quantities for these were recalculated using the scaling relations
and cosmology used in this work, and smaller symbols are used for QSOs with
redshifts $>$2.5. 
}
\label{figcontha}
\end{figure*}

The relationships between the three primary measurements used in this
section - \ha\ luminosity, \fwhmha, and continuum luminosity  -
are shown in Fig.~\ref{figcontha}.
The \ha\ and continuum luminosities are proportional (left panel of
Fig.~\ref{figcontha}),
with a continuum luminosity a factor of $\sim$2 higher than would be expected
from the extrapolation of the same relationship applied to
a large sample of SDSS galaxies \citep[log L$_{5100}\sim$43$-$45;][]{greho05}.
This correlation is unlikely to be due to a distance bias:  
the QSOs in the sample have similar redshifts, and 
a similar correlation is seen in the equivalent flux-flux plot. 
There is no clear relationship between \fwhmha\   
and both the luminosity of the \ha\ line and the continuum luminosity
(Fig.~\ref{figcontha}). 
Compared to the sub-mm detected
z$\sim$2 \textit{galaxies} with broad \ha\ or broad \hb\ from \citet{aleet08}, our
sample QSOs all have higher \ha\ luminosities, but show a considerable
overlap in \ha\ line widths.
In continuum luminosity and \fwhmha, our sample QSOs are very similar to the QSO
sample studied by \citet{sheet04}, but typically an order of magnitude more
luminous in the continuum than the QSO sample of \citet{netet07}.

Estimates for the black hole masses, Eddington fractions (i.e. the fractional
Eddington rate), mass accretion rates,
and FIR luminosities are listed in Table~\ref{figtab3} for the 10 sample QSOs.
We estimated the black hole masses and Eddington fractions of the sample QSOs 
from the empirical scalings derived from reverberation mapping results. 
Specifically, we used eqn. 1 of \citet{greho05} to convert \fwhmha\ 
to FWHM$_{\rm H\beta}$ and then estimated the black hole mass from 
FWHM$_{\rm H\beta}$ and L$_{5100}$ following 
eqn. 5 of \citet{vespet06}. 
Following \citet{kaset00}, \citet{maret04}, and \citet{nettra07} we use L$_{5100}$ = $\lambda$ L$_{\lambda}$ with
$\lambda$ = 5100\AA, and L$_{\rm Bol} = 7~\times$~L$_{5100}$. 
Eddington fractions were computed assuming this bolometric
luminosity, and the mass accretion rate was computed assuming a 10\% accretion 
efficiency ($\epsilon = 0.1$). 
We calculated the FIR luminosity from the 
sub-mm flux (\850\ or 1.2~mm) following 
\citet{mcmet99} and using $\beta = 1.5$ and dust temperature T$_{\rm d}$ = 40~K.
Finally, the star formation rate (SFR) was calculated from the FIR luminosity
following \citet{mcmet99} and using the \citet{ken98} relation, i.e. $\alpha=1.72$.

The relations we used are therefore:

$$\rm{FWHM}_{\rm H\beta}=(1.07 \pm
0.07)\times10^3\left(\frac{\rm{FWHM}_{\rm H\alpha}}{10^3
\rm{~km~s^{-1}}}\right)^{(1.03\pm0.03)} \rm{km~s^{-1}}$$

$$M_{BH}/M_{\odot}
=10^{6.91}\left(\frac{\rm{FWHM}_{\rm H\beta}}{1000\rm{~km~s^{-1}}}\right)^2\left(\frac{L_{5100}}{10^{44}\rm{~erg~s^{-1}}}\right)^{0.5}$$

l$_{\rm Edd}$~=~L$_{\rm Bol}$/L$_{\rm Edd}$~=~7 L$_{\rm 5100}$/L$_{\rm Edd}$ 

$$\frac{dm}{dt}=0.18~\frac{1}{\epsilon}\left(\frac{L_{\rm Bol}}{10^{46}\rm{~erg~s^{-1}}}\right)~\left(\rm{\frac{M_{\odot}}{yr^{-1}}
}\right)$$ 

$$M_d=\frac{S_{\nu}(\nu_{obs}D^2_L)}{\kappa_dB_{\nu}(\nu_{rest},T_d)(1+z)}$$

$$\kappa_d=\kappa_{850}\times\left(\frac{\nu_{rest}}{353~GHz}\right)^{\beta}$$

$$L_{FIR}\approx3.3\times\left(\frac{M_d}{10^8M_{\odot}}\right)\times10^{12}L_{\odot}$$

$$SFR=1.72\times10^{-10}\frac{L_{FIR}}{L_{\odot}}M_{\odot}~yr^{-1}$$

The lower rest-wavelength limit of our near-IR spectra is typically
5500\AA\ to 6000\AA.  Given that the spectral slope in
the 5500\AA--6500\AA\ range is relatively flat for the QSOs in our sample
we simply approximate that the 5100\AA\ flux is the same as the 5500\AA\ or
6000\AA\ flux.
\fwhmha, the FWHM of the \ha\ line profile as measured from the
sum of the double-Gaussian fit, was used in the black hole mass related calculations. 
If the FWHM of only the very broad (VBLR) Gaussian component of the
\ha\ line is used instead, the resulting black hole masses would increase
by a factor of $\sim$5.
We also estimated the black hole masses following the scaling
relations in \citet{maret08} with and without
radiation pressure effects, and in \citet{greho05};  in
all cases except one, the black hole mass estimates remain within a factor of 2--3 for
almost all QSOs in the sample. The exception is the \citet{maret08} relation which
includes radiation pressure effects: here the high continuum luminosities of the 
sample QSOs result in the radiation term dominating the black hole mass estimate,
i.e. the estimate is relatively independent of line width.
The use of \fwhmha\ and \Lha\ to estimate the black hole
mass \citep[following eqn. 6 of][]{greho05} leads to slightly lower
black hole masses in all QSOs (Table \ref{figtab3}) due to the slightly higher than usual 
luminosity of L$_{5100}$ compared to \Lha\ (Fig.~\ref{figcontha}).  

Errors in the black hole mass were estimated both
by Monte Carlo simulations assuming Gaussian errors for the input quantities 
and by the analytical calculation of the error propagation \citep{bevrob92}.
Both methods resulted in consistent error estimates, with errors around 10\%. 
Note, however, that these errors represent only the measurement errors and the
true error in the black hole mass is likely closer to a factor 4 \citep[e.g.,][]{vespet06}.
For the other derived quantities (e.g., l$_{\rm Edd}$) errors were calculated
by analytical error propagation.

Two of the sample QSOs have black hole estimates derived from the \mgii\ and \civ\
lines in \citet{sheet10}. 
For HS~B1049+4033, our estimated value lies close to the values (9.5 to 9.8 with
typical error of 0.07 as estimated from three different scaling relationships) estimated 
from \mgii\ and significantly lower than the value (10.21$\pm$0.06) estimated from \civ.
For HS~B1103+6416, our estimated value is significantly lower than that listed 
in \citet{sheet10}: 10.4$\pm$0.06 and 10.3$\pm$0.03 for \mgii\ and \civ, respectively. 
We remark that that using the FWHM of the very broad (VBLR) Gaussian component to 
calculate the \ha\ based black hole mass leads to a better agreement with the \civ\ 
and \mgii\ based black hole mass estimates.

\begin{figure*}[ht]
\includegraphics[bb=85 77 427 355,width=0.48\textwidth,clip]{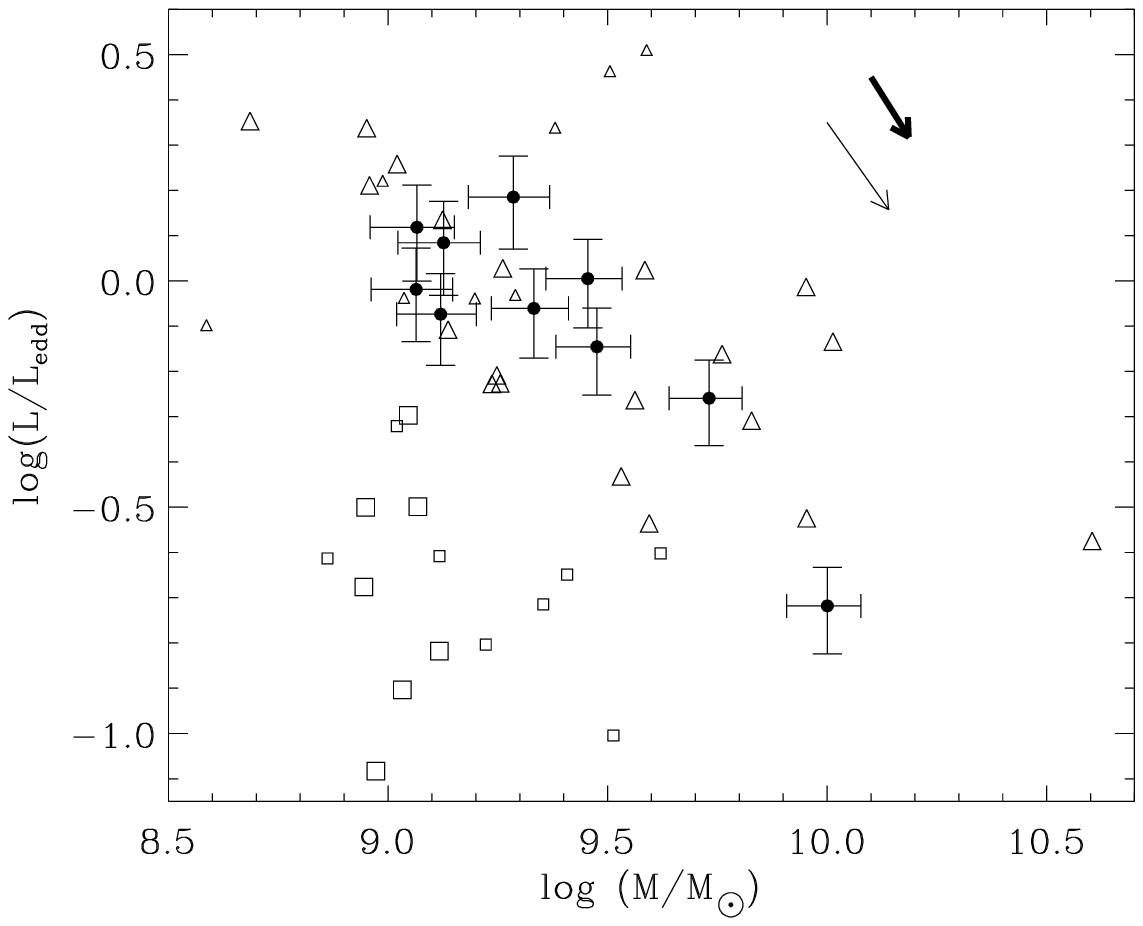}
\includegraphics[bb=85 77 427 355,width=0.48\textwidth,clip]{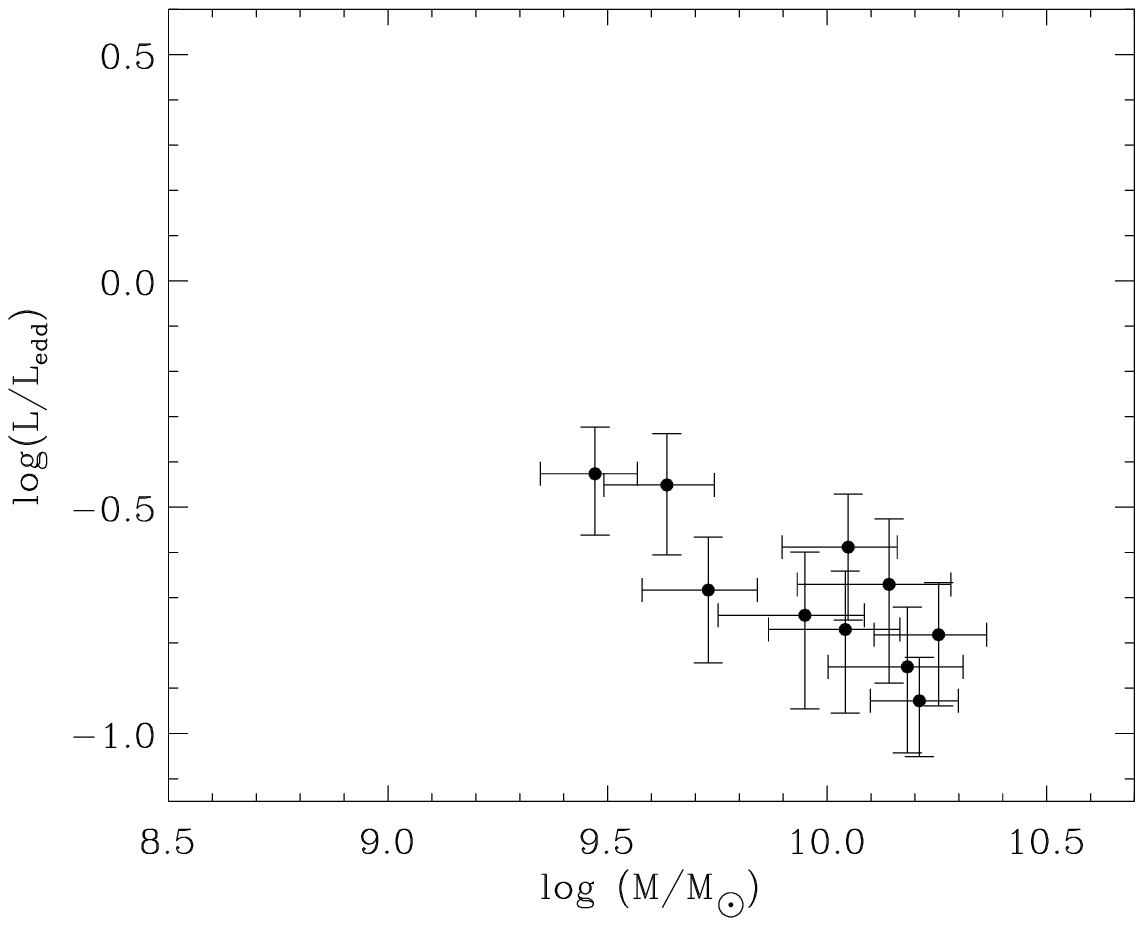}
\caption{\emph{Left:} The relationship between Eddington fraction and black hole 
mass contrasting QSOs in our sub-mm sample (filled circles plotted with their 1$\sigma$ errors)
to the QSO samples of \citet[][triangles]{sheet04} and \citet[][squares]{netet07}.
Quantities for the latter two samples were recalculated using the scaling relations
and cosmology used in this work, and smaller symbols are used for QSOs with
redshifts $>$2.5. The thin and thick arrows show the typical shift
of the \citet{sheet04} and \citet{netet07} points, respectively, if the QSO black hole 
masses are calculated via the equations used by these authors.
\emph{Right:} As in the left panel, but only for QSOs in our sub-mm sample and
in the case that only the very broad (VBLR) \ha\ component is used in the black
hole mass estimation. 
}
\label{figleddvsM}
\end{figure*}

The left panel of Fig.~\ref{figleddvsM} contrasts the relationship between
the
fractional Eddington rate and black hole mass of sub-mm detected QSOs in our sample 
to two other QSO samples at z$\sim$2; those of \citet{sheet04} and \citet{netet07}.  
All three samples have a similar range of black hole masses, though the QSOs of
the Shemmer et al. sample and our sample are accreting at relatively large Eddington
fractions. Combining this result with that seen in the right panel of Fig.~\ref{figcontha}
shows that our sub-mm detected QSO sample is virtually indistinguishable from
the QSO sample of \citet{sheet04}. This result is not very sensitive to the exact
scalings used in the black hole mass estimation (see the arrows in the figure):
if the scaling relations are modified, all QSOs in the figure
will shift their position in a relatively consistent manner. 
Note that the scaling relations we use imply that
L/L$_{\rm Edd} \propto$ L$_{\rm 5100}^{0.5}$ (FWHM$_{H\alpha})^{-2.06}$ and
M$_{\rm BH}  \propto$ L$_{\rm 5100}^{0.5}$ (FWHM$_{H\alpha})^{2.06}$. 
Thus Fig.~\ref{figleddvsM} plots highly related quantities, so that any subtle
correlations or anti-correlations seen are to be treated with caution.
As mentioned in Sect.~\ref{secintro}, it may be more correct to estimate 
QSO black hole masses from the FWHM of the very broad (VBLR) component instead 
of \fwhmha.
The right panel of Fig.~\ref{figleddvsM} demonstrates the implications of 
using the VBLR width in the black hole mass estimation: black hole masses
increase, and fractional Eddington rates decrease, by a factor of $\sim$5.
The comparison QSOs \citep{sheet04,netet07} will also change positions in
this plot, but in an unknown way as their VBLR widths are not listed by the
corresponding authors.

\begin{figure}[ht]
\includegraphics[bb=87 75 400 370,width=0.48\textwidth,clip]{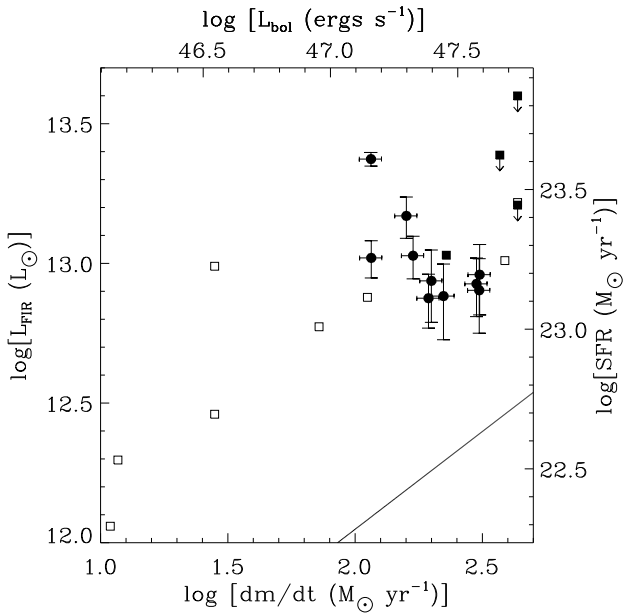}
\caption{The relationship between Far-Infrared (FIR) luminosity and the 
mass accretion rate onto the black hole for sub-mm detected QSOs (filled
circles with error bars). 
The equivalent axes of star formation rate (SFR) and 
bolometric luminosity are indicated on the right and top of the plot, respectively.
The sub-mm QSOs are close to the highest end of the relationship between the 
two quantities found to be valid over 6 orders of magnitude 
\citep[black line in the figure; ][]{tranet10}.
QSOs at z$\sim$2 observed in the mid-IR by \spitzer\ \citep{lutet08} are plotted as squares;
filled symbols are used for the QSOs in common with our sample.}
\label{figsfvsarate}
\end{figure}

The relationship between FIR luminosity (proportional to the star-formation
rate) and the mass accretion rate onto the black hole (proportional to L$_{\rm Bol}$) 
for the QSO sample is shown in Fig.~\ref{figsfvsarate}. 
\citet{tranet10} have studied a large sample of Seyferts and QSOs with
redshifts $0.2 \leq z \leq 3$, 
bolometric luminosities in the range $2\times10^{42} - 10^{48} \ergsec$,  
and SFRs in the range $10^{-1} - 10^{4}$ M\sun~yr$^{-1}$,
and find that these can be well characterized by a single relationship
(black line in Fig.~\ref{figsfvsarate}):
$$SFR~(M_{\odot}~yr^{-1}) \approx 32.8 \left(\frac{L_{Bol}~(erg/s)}{10^{46}}\right)^{0.7}$$ 
Our sample QSOs are found close to the highest end of this relationship
and offset from it in having SFRs almost one magnitude higher.
The sample of 12 millimeter-bright z$\sim$2 QSOs studied by \citet{lutet08}
with \spitzer\ are also shown for comparison: four of their targets are also
present in our sample. \citet{lutet08} estimated SFR and bolometric luminosity
from PAH-emission and 60\micron\ continuum, respectively, leading to
some differences from the values derived by us.


Finally, we compare the black hole mass distributions of the parent sample
of sub-mm detected and non-detected QSOs.
Of the 89 QSOs in the samples of \citet{priet03a} and \citet{omoet03},
40 appear in the compilation of \citet{sheet10}. \citet{sheet10} list,
among other quantities, the black hole masses derived from the \civ\ and \mgii\
lines and nearby continuum.
The black hole mass distributions of the sub-mm detected and sub-mm non-detected QSOs 
(right panel of Fig.~\ref{fighisto}) are not significantly different. To realize this figure, we derived
the black hole mass of the QSO by using, in order of preference, our \ha\ derived black 
hole mass, or the \mgii\ or \civ\ derived black hole mass listed in \citet{sheet10}.
This resulted in black hole masses for 14 sub-mm detections and 33 sub-mm non-detections.

\section{Discussion and Concluding Remarks}
\label{secdiscussion}

We have observed the \ha\ line and neighboring continuum in
10 of the 18 sub-mm detected QSOs from a parent sample of 83 
optically-bright QSOs at z$\sim$2. The properties of these 10 QSOs have
been compared with a sample of sub-mm detected galaxies
and two other samples of optically-bright \z2\ QSOs. 
We have also attempted to compare the 18 sub-mm detected QSOs with the 
65 sub-mm non-detected QSOs in the parent sample. This comparison is limited by two
factors; (a) the lack of accurate systemic velocities for the sub-mm 
non-detected QSOs and. (b) the relatively low sub-mm fluxes of some of the sub-mm detected QSOs
coupled with the relatively high detection limits of the sub-mm surveys
makes it difficult to clearly separate sub-mm bright QSOs from sub-mm dim 
QSOs. That is, it would not be surprising 
if some of the QSOs not detected in the sub-mm have sub-mm fluxes similar to the weakest
sub-mm detected QSOs. Nevertheless, the parent sample used here (18 detections in a 
sample of 83 QSOs) is sufficiently large to obtain meaningful comparisons 
between sub-mm-bright QSOs and other QSOs.

The \ha\ line profiles can be decomposed into contributions from
both a very broad line region (VBLR) and an intermediate-width line region (ILR), 
similar to the cases of \hb\ and \civ\ in QSOs. 
Both components show similar redshifts in most of the QSOs, and the ratios of the VBLR to
ILR widths (1.7--5 with a median of 3.0) are not different from the typical ratios 
($\sim$2.5) found for \hb\ profiles in QSOs \citep{huet08}. 
The ILR has been interpreted to be the outer portion of the BLR and may originate from a
physically distinct region \citep{huet08}. Its inclusion in the black hole
mass estimation has been questioned \citep{zhuet09}. 
Following previous works, we estimate the QSO black hole masses using
the combined VBLR+ILR width, but also discuss the implications of using 
only the VBLR width.

We have made the reasonable assumption that the intermediate-width (ILR) 
component of the broad \ha\ emission closely traces the true systemic velocity of the QSO.
A blind comparison of \ha-based redshifts derived here with previous
redshifts based on high-ionization rest-frame UV-lines
shows larger than expected redshift differences (Fig.~\ref{figzhbqs}).
However, the largest differences are seen in QSOs whose rest-frame UV
redshifts are imprecise, inaccurate, and/or for which redshift errors have
not been provided. For example, the HBQS and LBQS quote redshifts 
for these QSOs to the second decimal place (i.e. an accuracy of $\sim$1500~\kms\ 
if the redshift values were rounded or $\sim$3000~\kms\ if truncated).
Limiting ourselves to the QSOs with accurate rest-frame UV redshifts,
the typical rest-frame optical to rest-frame UV redshift
difference is $\sim$0.013, or $\sim$4000\kms. 
Further limiting ourselves to only sample QSOs with reliable rest-frame UV spectroscopy
from SDSS, we find larger than average velocity offsets (up to 3400\kms) for 
individual rest-frame UV lines (Fig.~\ref{figzsdss}). 
Given the large obscuration expected in sub-mm QSOs (nuclear dust
would obscure gas on the far side of the AGN), these velocity offsets most likely trace 
ionized gas outflows.
The lack of accurate redshifts for the sub-mm non-detected QSOs makes it impossible to
definitively test whether these powerful inflows/outflows are found
only in the sub-mm detected QSOs and not in the sub-mm non-detected QSOs
from the parent samples.

Black hole mass estimates used in this work are based on the FWHM of
the \ha\ line (\fwhmha) and the continuum luminosity. 
Using the \ha\ luminosity instead of continuum luminosity lowers the masses
by $\sim$3. Using the width of the very broad Gaussian component 
(i.e. the VBLR only instead of both VBLR and ILR) to estimate the black hole
mass increases the black hole mass estimates by $\sim$5
and changes the interpretation of the black hole growth phase in the QSO sample
(Fig.~\ref{figleddvsM}).
This uncertainty, coupled with the uncertainty of applying reverberation-mapping-based
scaling relations at these high continuum luminosities limits the reliability of our results,
and those of other similar works.

We find no significant differences in the distributions of black hole masses 
or emission-line velocity offsets between the subsamples of sub-mm detected
and non-detected QSOs from the parent samples of \citet{priet03a}
and \citet{omoet03}. 
Furthermore, \citet{priet03a} and \citet{omoet03} 
did not find any other predictor of sub-mm loudness. The lack of 
rest-frame optical spectroscopy of the sub-mm non-detected QSOs limits
any comparison, though it is reasonable to assume that their properties
are similar to other z$\sim$2 QSOs (e.g. the samples of \citet{sheet04}
and \citet{netet07}).

The sub-mm detected QSOs studied here have the highest rest-frame optical continuum 
luminosities among sub-mm detected galaxies thus far studied in rest-frame optical 
spectroscopy (e.g. Fig.~\ref{figcontha}).
These sub-mm detected QSOs host black holes in the typical range 
of luminous QSOs at redshift z$\sim$2, and are accreting at close to their
Eddington limit (Fig.~\ref{figleddvsM}). In black hole mass and
Eddington fraction they are indistinguishable from the optically-luminous QSO sample of 
\citet{sheet04}. The latter sample has not been studied in the sub-mm though
it is reasonable to assume that it is dominated by sub-mm dim QSOs.
If black hole masses in
the sub-mm detected QSOs are estimated from the VBLR width only, then black hole
masses increase, and Eddington fractions decrease, by a factor of $\sim$5. In this
case the sub-mm detected QSOs host relatively massive black holes which are accreting at a fraction of
the Eddington rate. In this case we are unable to make a meaningful comparison to the QSO samples of 
\citet{sheet04} and \citet{netet07} as VBLR widths are not quoted in their studies. 
In brief, the sub-mm detected QSOs are, thus far, indistinguishable from the
typical z$\sim$2 QSO population apart from their relatively high sub-mm
fluxes.

\citet{netet07} and \citet{tranet10} have demonstrated a correlation
between SFR and mass accretion rate in active galaxies, valid over six orders of magnitude. 
The extreme SFRs and mass accretion rates of 
the sub-mm detected QSOs places them in the vicinity of the extreme end of this trend
(Fig.~\ref{figsfvsarate}).
However, they are offset from the relationship in having a factor $\sim$10 higher SFR. 
Within the sub-mm detected QSO sample, there
is a weak indication that the strongest star forming QSOs have the lowest 
black hole masses and mass accretion rates (Fig.~\ref{figsfvsarate}), but this
weak trend is within the scatter seen when other QSOs are 
included in the plot.
\citet{maiet07} and \citet{lutet08} have studied the SFR accretion-rate correlation in a sample of optically-luminous 
QSOs observed with Spitzer and noted that the SFR potentially flattens at the highest bolometric luminosities.
Adding the data from our sample of QSOs (which changes some of the previous
large upper-limits on the SFR into known lower values) appears to support their findings.

\begin{acknowledgements}
We note with great sadness that Robert Priddey passed away on 20th February 2010.
The authors acknowledge the contribution of the full TNG/NICS engineering team,
and  thank B. Trakhtenbrot, H. Netzer, and the anonymous referee, for useful suggestions.
KGI acknowledges the support of a PPARC fellowship during the time at which
this project was initiated. 
NN acknowledges funding from Conicyt-ALMA 31070013 and 3108022, the Fondap Center
for Astrophysics, BASAL PFB-06/2007, and Fondecyt 1080324. 
Part of this work was carried out while NN was a postdoctoral
fellow at Arcetri Observatory and Kapteyn Institute.

\end{acknowledgements}

 \begin{table}
  \dummytable\label{figtab1}
 \end{table}
 \begin{table}
  \dummytable\label{figtab2}
 \end{table}

\begin{table}
 \dummytable\label{figtab3}
\end{table}


\begin{figure*}[ht]
\includegraphics[angle=0,bb=0 0 612 792,width=\textwidth,clip]{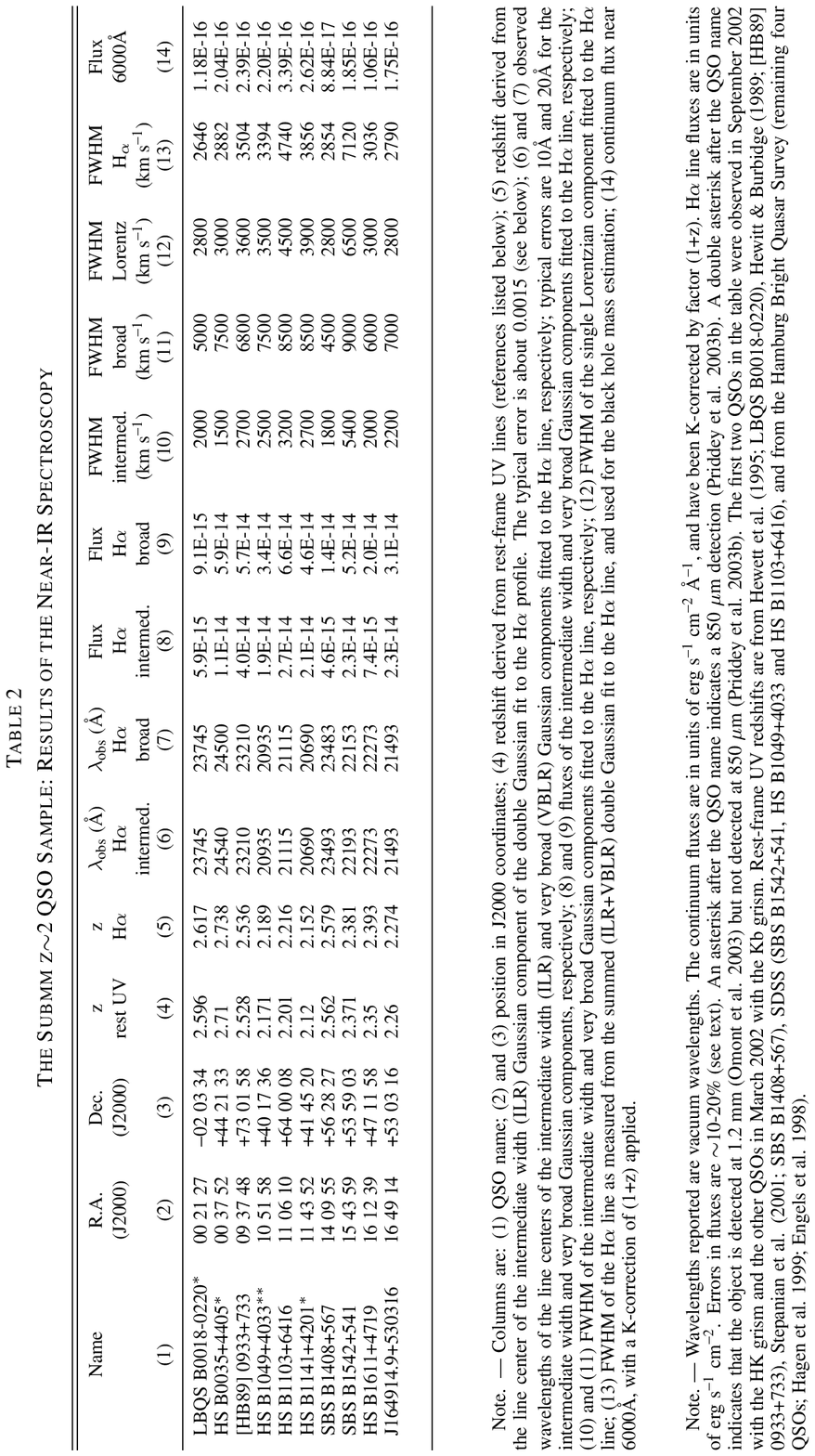}
\end{figure*}

\clearpage

\begin{figure*}[ht]
\includegraphics[bb=0 150 612 655,width=\textwidth,clip]{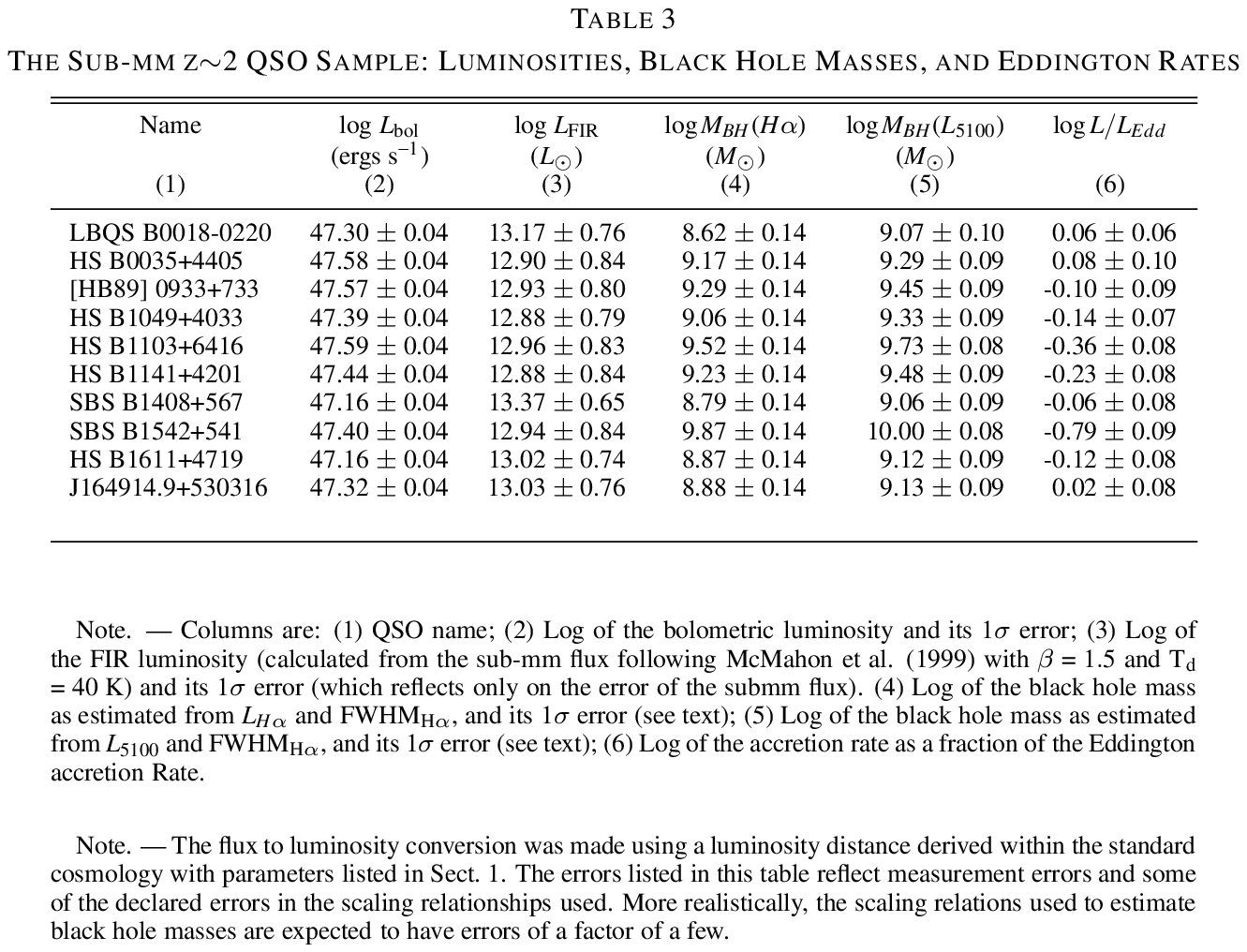}
\end{figure*}

\end{document}